\documentclass[aps,english,twocolumn]{revtex4}
\usepackage{mathrsfs}
\usepackage{graphicx}
\usepackage{amsmath, amssymb}
\usepackage{babel}

\begin{document}

\title{Coherent state path integral approach to correlated electron systems with deformed Hubbard operators: from Fermi liquid to Mott insulator}
\preprint{1}

\author{Xiao-Yong Feng}
\email[fxyong@hznu.edu.cn]{}
\author {Jianhui Dai}
\email[daijh@hznu.edu.cn]{}
\affiliation{Department of Physics, Hangzhou Normal University, Hangzhou 310036, China}
\date{April 27, 2019}

\begin{abstract}
 In strongly correlated electron systems the constraint which prohibits the double electron occupation at local sites can be realized by either the infinite Coulomb interaction or the correlated hopping interaction described by the Hubbard operators, but they both render the conventional field theory inapplicable. Relaxing such the constraint leads to a class of correlated hopping models based on the deformed Hubbard operators which smoothly interpolate the locally free and strong coupling limits by a tunable interaction parameter $0\leq \lambda\leq 1$. Here we propose a coherent state path integral approach appropriate to the deformed Hubbard operators for {\it arbitrary} $\lambda$. It is shown that this model system exhibits the correlated Fermi liquid behavior characterized by the enhanced Wilson ratio for all $\lambda$. It is further found that in the presence of on-site Coulomb interaction a finite Mott gap appears between the upper and lower Hubbard bands, with the upper band spectral weight being heavily reduced by $\lambda$. Our approach stands in general spatial dimensions and reveals an unexpected interplay between the correlated hopping and the Coulomb repulsion.
\end{abstract}

\maketitle

A major challenge in interacting electron systems comes from the lack of proper theoretical approaches for various kinds of many body correlations which are ubiquitous and variable in solids.  The electron correlations are of fundamental importance in the formation of novel quantum phases in the condensed matter physics\cite{Fulde-book91,Coleman-rev03}. They can be captured by the short-ranged Coulomb interaction and the correlated hopping (CH) interaction as described in the Hubbard model\cite{Hubbard-prs63,Hubbard-prs64}or its variants\cite{Foglio-prb79,Emery-prl87,Simon-prb93,Gehhard-book97} as for the electron systems with narrow bandwidths\cite{Imada-rmp98}. For the free hopping system with a weak Coulomb interaction the Landau's Fermi liquid\cite{Landau-56} develops continuously from the free electron gas. An opposite extreme case is the strong coupling limit when the Coulomb interaction tends to infinity so that the perturbation treatment around the non-interacting limit is invalid. In this case the Fermi liquid scenario might either persist but with strong correlation effect or breakdown, driving the half-filled system to an insulator \cite{Mott-prs49,Hubbard-prs64}. The interaction-driven many-body features are relevant to a number of unconventional quantum phenomena ranging from Mott transition\cite{Mott-prs49,Gehhard-book97}, Peierls dimerization\cite{Kivelson-prl87}, high temperature superconductivity\cite{Hirsch-physicac89,Lee-rmp06}, heavy fermion physics \cite{Stewart-rmp01,Si-science10}, to quantum magnetism\cite{Loehneysen-rmp07,Zhou-rmp18}.

Theoretically, the main difficulty in tackling the strong coupling limit comes from the constraint that the possibility of the double occupation at the same lattice site for two electrons with opposite spin polarizations should be completely excluded. This no-double-occupation constraint can be imposed by applying the Gutzwiller projection operator ${\cal P}_G =\prod_{j} (1-n_{j\uparrow}n_{j\downarrow})$ \cite{Gutzwiller-prl63} on a given Hamiltonian ${\hat H}$ so that the resulting model system ${\tilde {\hat H}}={\cal P}_G{\hat H}{\cal P}_G$ is defined on the truncated Hilbert space without any double occupation states.
The truncation of system's Hilbert space such as in the t-J model\cite{Zhang-prb88}severely prohibits applications of the conventional many-body techniques including the path integral and diagram expansion. The slave-particle method is then devised in order to implement the constraint, but at the cost of introducing additional slaved gauge degrees of freedom\cite{Kotliar-prl86}. Another approach is to apply the local Hubbard X operators in a concise way \cite{Hubbard-prs65,X2}, among which the operators $X_i^{0\sigma}=c_{i\sigma}(1-n_{i\bar{\sigma}})$ obviously exclude the double occupation at local sites. However, the Hubbard X operators possess no conventional anti-commutation rules, leading to a cumbersome Lagrangian \cite{X3} and complicated X-operator-based diagram technique \cite{Ovchinikov-book04}.

Given the fact that the Coulomb interaction is actually not infinite, the many-body correlation effect in realistic materials is frequently investigated by the variational Gutzwiller projection method using the partial projection operator ${\cal P}_G(\lambda) =\prod_{j} (1-\lambda n_{j\uparrow}n_{j\downarrow})$ with $0\leq \lambda\leq 1 $\cite{Gutzwiller-pra65,Brinkmann-prb70}. Treating $\lambda$ as a variational parameter can optimize the effect of the Coulomb $U$ which drives the Mott transition at half filling\cite{Brinkmann-prb70,Izyumov-rev95}.
The Gutzwiller variational approach combined with the density functional theory has been shown powerful in understanding the many-body correlation effect in 3d or 4f electronic materials\cite{Daixi-epl08,Ho-prb08,Daixi-prb09,Kotliar-prl14}.

These developments naturally stimulate the research interest in a class of correlated electron systems whose Hamiltonians are constructed in terms of the deformed Hubbard X operators $\tilde{X}_i^{0\sigma}=c_{i\sigma}(1-\lambda n_{i\bar{\sigma}})$, the operators which smoothly interpolate the conventional electron operators ($\lambda=0$) and the Hubbard operators ($\lambda=1$)\cite{Chen-prb02}. The deformation induces a complicated CH motion involving four- and six-fermions interacting terms in the kinetic part of the original Hamiltonian when $\lambda \neq 0$. This is what the CH interaction means in the present study. More generally, CH interaction emerges rather naturally in the construction of tight-binding Hamiltonians involving variable electron-electron or electron-phonon interactions\cite{Hubbard-prs63,Hubbard-prs64,Foglio-prb79,Kivelson-prl87,Hirsch-physicac89,Simon-prb93}. In one-dimensional systems, such CH interaction has been examined by using the bosonization technique and renormalization group for small $\lambda$ \cite{Chen-prb02,Japaridze-prb99,Arrachea-prb99,Ma-mpla07}. For $\lambda$ close to unit these analytical methods are inapplicable again. In spite of this fact, the limit $\lambda=1$ has been long expected equivalent to the case of infinite Coulomb interaction imposed by the Gutzwiller projection although the effect caused by CH interaction has not been fully examined so far.

In order to clarify the basic physics of CH interaction and its interplay with Coulomb interaction in general dimensions, it is highly desirable to study the correlated hopping models (CHMs) for arbitrary $\lambda$.
In the present paper, we shall develop a new path integral approach for these models using the coherent state representation of the  deformed Hubbard $\tilde{X}^{0\sigma}$ operators. We will show how the path integral in this representation can provide interesting results for the thermodynamics property in arbitrary spatial dimensions.

To begin with, let us reiterate the definition of the deformed Hubbord $X^{0\sigma}$ operators \cite{Hubbard-prs65,Chen-prb02}
\begin{equation}
\tilde{X}_i^{0\sigma}=c_{i\sigma}(1-\lambda n_{i\bar{\sigma}}),
\end{equation}
where $c_{i\sigma}$ is the annihilation operator for the electron with spin $\sigma$(=$\uparrow$ or $\downarrow$) at site $i$ and $n_{i\bar{\sigma}}=c^{\dag}_{i\bar{\sigma}}c_{i\bar{\sigma}}$ is the particle number operator at the same site with opposite spin, $0\leq \lambda\leq 1$.
Parallel to the path integral representation for the conventional electron operators\cite{Schulman-book81,Zhang-rmp90}, we seek for the coherent states  $|\xi_i\rangle $ defined as the eigenstates of the deformed Hubbard operators
\begin{equation}
\tilde{X}_i^{0\sigma}|\xi_i\rangle=\xi_{i\sigma}|\xi_i\rangle,
\end{equation}
with the eigenvalues $\xi_{i\sigma}$ being Grassmann numbers. Note that $\tilde{X}_i^{0,\uparrow}$ and $\tilde{X}_i^{0,\downarrow}$ have the common eigenstates because they are anti-commutative to each other. We find that these eigenstates can be constructed in the form
\begin{equation}
|\xi_{i}\rangle = \exp{\left[-\left(\sum_{\sigma}\xi_{i\sigma}c^{\dag}_{i\sigma}+\frac{\lambda}{1-\lambda}\xi_{i\uparrow}\xi_{i\downarrow}c^{\dag}_{i\uparrow}c^{\dag}_{i\downarrow}\right)\right]}|0\rangle
\end{equation}
where $|0\rangle$ is the null state of the original annihilation operators $c_{i\sigma}$.
The coherent states are well-defined for all $\lambda$ except for the singular point $\lambda=1$, nevertheless, the limit of $\lambda\rightarrow1$ still exits in the final results. 

It is easy to verify that the overlap of any two coherent states is given by
\begin{eqnarray}
\langle\xi_i|\xi'_i\rangle=\exp{\left[\sum_{\sigma}\xi^*_{i\sigma}\xi'_{i\sigma}+(u-1)\xi^*_{i\downarrow}\xi^*_{i\uparrow}\xi'_{i\uparrow}\xi'_{i\downarrow}\right]}
\end{eqnarray}
with $u=(1-\lambda)^{-2}$. The set of all these coherent states is overcomplete, satisfying the following relationship
\begin{eqnarray}\label{completeness}
\frac{1}{u}\int\prod_{\sigma}d\xi^*_{i\sigma}d\xi_{i\sigma}e^{-u\sum\limits_{\sigma}\xi^*_{i\sigma}\xi_{i\sigma}-(u^2-u)\prod\limits_{\sigma}\xi^*_{i\sigma}\xi_{i\sigma}}|\xi_i\rangle\langle\xi_i|=1.
\end{eqnarray}

Now we consider a Hamiltonian ${\hat H}(\{\tilde {X}^{\dagger}\}, \{\tilde {X}\})$, expressed in terms of the deformed X operators ( $\{\tilde{X }\}$ indicates the set of all $\tilde{X}^{0\sigma}_i$). In the thermal equilibrium state at temperature $T$, the partition function is $ Z=tr e^{-\beta\hat{H}}$, $\beta=1/{k_B T}$. Dividing the (imaginary) time interval ($\beta$) by $M$ number of slices and evaluating the trace in the representation of the coherent states\cite{Schulman-book81,Zhang-rmp90}, the partition function is expressed as the following path-integral\cite{note1}
\begin{equation}\label{partition}
Z = \int D(\xi^*,\xi)e^{-S(\xi^*,\xi)}.
\end{equation}
Where, the classical action takes the form
\begin{eqnarray}
S &=&\sum_{i,m,\sigma}(u\tilde{n}_{i\sigma,m}-\xi^*_{i\sigma,m}\xi_{i\sigma,m-1})\nonumber\\
&+&\sum_{i,m}(u^2-u)\prod_{\sigma}\tilde{n}_{i\sigma,m}-\sum_{i,m}(u-1)\prod_{\sigma}\xi^*_{i\sigma,m}\xi_{i\sigma,m-1}\nonumber\\
&+&\sum_{m}\epsilon H[\{\xi^*_{m}\},\{\xi_{m-1}\}],
\end{eqnarray}
$D(\xi^*,\xi)= \lim\limits_{M\rightarrow\infty}\prod\limits_{i,m}(\frac{1}{u}) d\xi^*_{i\uparrow,m}d\xi_{i\uparrow,m}d\xi^*_{i\downarrow,m}d\xi_{i\downarrow,m}$ is the measure,
$\xi_{i\sigma,m}$ the Grassmann eigenvalues of the coherent states $|\xi_{i}\rangle$ in the $m$-th slice,
$\tilde{n}_{i\sigma,m}=\xi^*_{i\sigma,m}\xi_{i\sigma,m}$, $H[\{\xi^*_{m}\},\{\xi_{m-1}\}]$ the corresponding classical Hamiltonian, and $\epsilon=\beta/M$ the width of the time slices.

The obtained classical action looks absurd: there are additional four-fermions interaction terms (on the second line of Eq.(7)) which are seemingly divergent at the limit $M\rightarrow \infty$, hindering the direct application of the continuous field theory. To overcome this difficulty, we shall calculate the partition function by the Grassmann integration over the discrete time slices. Here the key observation is that the contribution from the classical action is actually measured by the pre-factor $1/u$ which eventually guarrantees the finiteness of the problem.

\begin{figure}[ht]
\includegraphics [width=7cm]{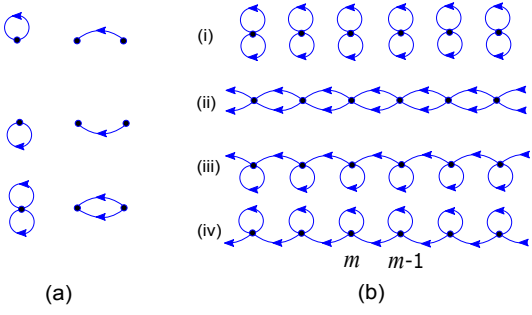}
 \caption{(a)The building blocks. (b)The contributing diagrams for the toy model.} \label{fig1}
\end{figure}

In order to show the validity of our approach, we first consider a simple toy model which contains the homogeneous on-site quadratic term:
\begin{eqnarray}\label{Hw}
\hat{H}_{w} = -\sum_{i\sigma}w_{\sigma}(\tilde{X}_i^{0\sigma})^{\dag}\tilde{X}_i^{0\sigma}.
\end{eqnarray}
The corresponding action contributed from $\hat{H}_{w}$ is
$S_{w}=-\sum_{i\sigma,m}\epsilon w_{\sigma}\xi^*_{i\sigma,m}\xi_{i\sigma,m-1}$.
Owing to the site-independence of the present case, we just focus on a given site $i$ and expand the exponential in the partition function Eq.(\ref{partition}) to
\begin{eqnarray}\label{expand1}
\prod_{m}&&\left(1-\sum_{\sigma}f_{1\sigma}\tilde{n}_{i\sigma,m}+f_2\prod_{\sigma}\tilde{n}_{i\sigma,m}\right)\nonumber\\
&&\left(1+\sum_{\sigma}f_{3\sigma}\xi^*_{i\sigma,m}\xi_{i\sigma,m-1}+f_4\prod_{\sigma}\xi^*_{i\sigma,m}\xi_{i\sigma,m-1}\right)
\end{eqnarray}
with $f_{1\sigma}=f_2=u$, $f_{3\sigma}=1+\epsilon w_{\sigma}$, and $f_4=u+\epsilon\sum_{\sigma}w_{\sigma}$.
All possible contributions to the partition function can be schematically summarized in Fig.\ref{fig1} where the dots represent the positions of time slices; the lines with arrows coming from or to the slice $m$ represent $\xi_{i\sigma,m}$ and $\xi^*_{i\sigma,m}$, respectively (the lines above/below the dots are for the spin up/down components).

Fig.\ref{fig1} (a) represents the various terms in (\ref{expand1}) which constitute the building blocks in the path integration. When all the dots are connected by a pair of out- and in-lines for each spin components, the corresponding Grassmann integration is non-vanishing. We call such a configuration the contributing diagram. There are four distinct contributing diagrams as shown in the Fig.\ref{fig1} (b), with the corresponding contributions being given by
\begin{eqnarray}
  &(i)&:\lim\limits_{M\rightarrow\infty}\frac{1}{u^{M}}f_2^M=1,\nonumber\\
  &(ii)&:\lim\limits_{M\rightarrow\infty}\frac{1}{u^{M}}f_4^M=e^{\beta\sum\limits_{\sigma}\frac{w_{\sigma}}{u}},\nonumber\\
  &(iii)&:\lim\limits_{M\rightarrow\infty}\frac{1}{u^{M}}(f_{1\downarrow}f_{3\uparrow})^M=e^{\beta w_{\uparrow}},\nonumber\\
  &(iv)&:\lim\limits_{M\rightarrow\infty}\frac{1}{u^{M}}(f_{1\uparrow}f_{3\downarrow})^M=e^{\beta w_{\downarrow}}.\nonumber
\end{eqnarray}
Therefore, we reproduce the desired exact result
\begin{eqnarray}\label{Z1}
Z_{w}=1+e^{\beta\sum\limits_{\sigma}\frac{w_{\sigma}}{u}}+\sum_{\sigma}e^{\beta w_{\sigma}}.
\end{eqnarray}

\begin{figure}[ht]
\includegraphics [width=5cm]{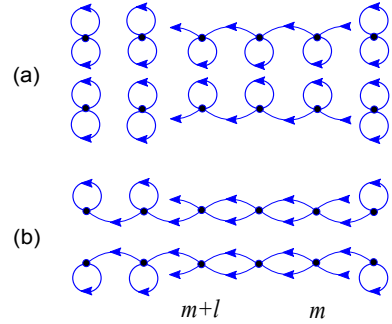}
 \caption{Two types of contributing diagrams generated by the $\epsilon^2$ terms.} \label{fig2}
\end{figure}

Now, we consider a non-trivial CHM: $\hat{H}=\hat{H}_{t} + \hat{H}_{w}$,  where $\hat{H}_{w}$ is given by (\ref{Hw}), and
\begin{eqnarray}\label{H}
\hat{H}_{t}=-\sum_{\langle ij\rangle\sigma}t\left[(\tilde{X}^{0\sigma}_i)^{\dag}\tilde{X}^{0\sigma}_{j} + h.c.\right],
\end{eqnarray}
$\langle ij \rangle$ indicates the nearest neighbor sites. The corresponding action is
$S_t=-\sum_{\langle ij\rangle\sigma,m}\epsilon t(\xi^*_{i\sigma,m}\xi_{j\sigma,m-1}+\xi^*_{j\sigma,m}\xi_{i\sigma,m-1})$.
Performing the Fourier transformation
$\xi_{i\sigma,m}=\frac{1}{\sqrt{N\beta}}\sum_{k}\xi_{k\sigma}e^{-i\omega_n\tau_m+i\vec{k}\cdot \vec{r}_i}$,
$k=(\omega_n,\vec{k})$, $\omega_n = (2n-1)\pi/\beta$ ($n=1,2,\cdots, M$), and $\tau_m=m\epsilon$, we have
$S_t= \sum_{k\sigma}\varepsilon_{k}\xi^*_{k\sigma}\xi_{k\sigma}$,
where $\varepsilon_{k}=\varepsilon_{\vec{k}}e^{i\omega_n\epsilon}$ and $\varepsilon_{\vec{k}}$ is the energy band of free fermions. This contribution can be combined with that of $\hat{H}_{w}$ if we introduce a pair of auxiliary Grassmann fields ($\eta_{k\sigma}, \eta^*_{k\sigma}$) by using the identity:
\begin{eqnarray}\label{St}
&\exp{(-S_t)}&=\prod_{k\sigma}(-\varepsilon_{k})\int \prod_{k\sigma}(d\eta^*_{k\sigma}d\eta_{k\sigma})\\
&&\exp{\sum_{k\sigma}(\varepsilon^{-1}_{k}\eta_{k\sigma}^*\eta_{k\sigma}+\eta^*_{k\sigma}\xi_{k\sigma}+\xi_{k\sigma}^*\eta_{k\sigma})}.\nonumber
\end{eqnarray}
Fourier transformed back to the real lattice space, $\sum_{k\sigma}(\eta^*_{k\sigma}\xi_{k\sigma}+\xi_{k\sigma}^*\eta_{k\sigma})$ becomes  $\epsilon\sum_{i\sigma}(\eta^*_{i\sigma}\xi_{i\sigma}+\xi_{i\sigma}^*\eta_{i\sigma})$. Then, expanding the exponential we find that the contributions to the partition function come only from the even power terms of $\epsilon$. The $\epsilon^2$ terms generate two new types of contributing diagrams as illustrated in Fig. \ref{fig2}, with the following contributions
\begin{eqnarray}
&(v)&:\lim\limits_{M\rightarrow\infty}\frac{\epsilon^2}{u^M}\sum_{\sigma,l=0}^{M-1}\sum_{m=1}^M f_{1\bar{\sigma}}^{l+1}f_2^{M-l-1}f_{3\sigma}^{l}\eta^*_{i\sigma,m+l}\eta_{i\sigma,m},\nonumber\\%\xi_{i\sigma,m+\delta}\xi_{i\sigma,m}^*
&(vi)&:\lim\limits_{M\rightarrow\infty}\frac{\epsilon^2}{u^M}\sum_{\sigma,l=0}^{M-1}\sum_{m=1}^Mf_{1\sigma}^{M-l-1}f_{3\bar{\sigma}}^{M-l}f_4^{l}\eta^*_{i\sigma,m+l}\eta_{i\sigma,m}.\nonumber
\end{eqnarray}
In the frequency space, the summation of the above terms gives rise to $-\sum\limits_{\omega_n,\sigma}g_{\sigma}(\omega_n)\eta^*_{i\sigma}(\omega_n)\eta_{i\sigma}(\omega_n)$, with
\begin{eqnarray}\label{g}
g_{\sigma}(\omega_n)=\frac{1+e^{\beta w_{\sigma}}}{i\omega_n+w_{\sigma}}+\frac{\frac{1}{u}\left(e^{\beta w_{\bar{\sigma}}}+e^{\beta\sum_{\sigma'}\frac{w_{\sigma'}}{u}}\right)}{i\omega_n+\sum_{\sigma'}\frac{w_{\sigma'}}{u}-w_{\bar{\sigma}}}.
\end{eqnarray}

Integrating out the $\xi$ fields, the partition function is
\begin{eqnarray}\label{Z2}
Z& = &\prod_{k\sigma}(-\varepsilon_k)\int \prod_{k\sigma}(d\eta^*_{k\sigma}d\eta_{k\sigma})\exp{\sum_{k\sigma}(\varepsilon^{-1}_{k}\eta_{k\sigma}^*\eta_{k\sigma})}\nonumber\\
&& \prod_i\left(Z_w-\sum_{\omega_n,\sigma}g_{\sigma}(\omega_n)\eta^*_{i\sigma}(\omega_n)\eta_{i\sigma}(\omega_n)+\cdots\right),
\end{eqnarray}
where $Z_w$ is given by Eq. (\ref{Z1}) and the "$\cdots$" represents the contribution from the high order terms of $\epsilon$.

So far our approach is rigorous if the high order terms of $\epsilon$ are included.
For the present purpose, we are interested in a closed analytical expression of $Z$, which can be obtained by using the approximation:
$Z_w-\sum_{\omega_n,\sigma}g_{\sigma}(\omega_n)\eta^*_{i\sigma}(\omega_n)\eta_{i\sigma}(\omega_n)+\cdots
\approx Z_w\prod_{\omega_n,\sigma}\left[1-Z_w^{-1}g_{\sigma}(\omega_n)\eta^*_{i\sigma}(\omega_n)\eta_{i\sigma}(\omega_n)\right]$.
This is understood as a re-summation in the Eq.(\ref{Z2}) under the ladder approximation. Integrating out the $\eta$ fields, we finally obtain the following expression
\begin{eqnarray}\label{Hc}
Z = Z_w^N\prod_{k\sigma}\left[1-Z_w^{-1}g_{\sigma}(\omega_n)\varepsilon_{\vec{k}}\right].
\end{eqnarray}

\begin{figure}[ht]
\includegraphics [width=8.5cm]{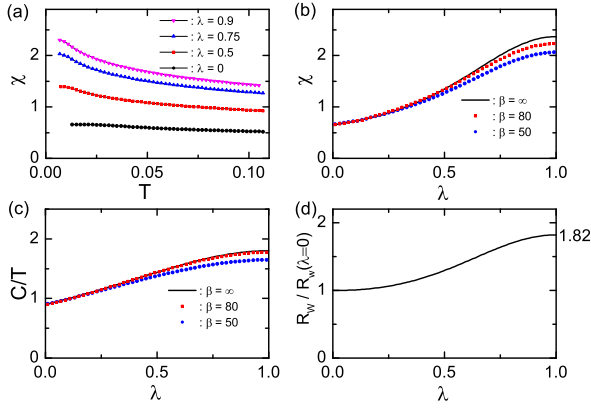}
 \caption{The susceptibility (a,b), specific heat coefficient(c), and Wilson ratio(d) for the CHM. } \label{fig3}
\end{figure}

Interestingly, when $u=1$, the partition function returns to the exact result for the free hopping model (the case with $\lambda=0$), demonstrating the validity of the above approximation. Using Eq.(15), the low temperature specific heat for $w_{\sigma} = 0$ is obtained as
$ C =\frac{4}{1+\frac{1}{u}}N(0)k_B^2T $,
where $N(0)$ is the density of states for the corresponding free hopping electrons around $\varepsilon_{\vec{k}}=0$.
While, the susceptibility at zero temperature is obtained as
$\chi = \left[\frac{4}{1+\frac{1}{u}}+\frac{\pi^2}{3}\frac{(1-\frac{1}{u})^2}{(1+\frac{1}{u})^3}\right] N(0)
$.
Figs. \ref{fig3}(a-c) show the $\lambda$-dependence of the susceptibility and specific heat coefficient at several temperatures using the square lattice dispersion $\varepsilon_{\vec{k}}= - 2t (\cos k_x + \cos k_y)$ and $t=1$ (see more details in Appendices E and F). These results show that the CHM displays the Fermi liquid behavior with the renormalized specific heat coefficient and Pauli susceptibility, both enhanced by $\lambda$. In particular, the Wilson ratio $R_W=\chi : C/T$ is enhanced to $R_W=1.82$ for $\lambda=1$ as shown in Fig. \ref{fig3}(d) (where the Wilson ratio for $\lambda=0$ is scaled to unit). The nearness of this value to the one for the magnetic instability ($R_W=2$) indicates the strong correlation effect in this Fermi liquid phase.

Finally, we consider the Hubbard model in the presence of CH interaction, described by the Hamiltonian
\begin{eqnarray}
\hat{H}=\hat{H}_t-\sum_{i\sigma}\mu_{\sigma} n_{i\sigma}+\sum_{i}Un_{i\uparrow}n_{i\downarrow},
\end{eqnarray}
where $\mu_{\sigma}$ is the spin-dependent chemical potential and $U$ the on-site Coulomb repulsion. It is remarkable that these two terms can be reexpressed in terms of the deformed Hubbard operators as
\begin{eqnarray}\label{U}
&-&\sum_{\sigma}\mu_{\sigma}(\tilde{X}_i^{0\sigma})^{\dag}\tilde{X}_i^{0\sigma}\nonumber\\
&+& \left[uU-(u-1)\sum_{\sigma}\mu_{\sigma}\right](\tilde{X}_i^{0\downarrow})^{\dag}(\tilde{X}_i^{0\uparrow})^{\dag}\tilde{X}_i^{0\uparrow}\tilde{X}_i^{0\downarrow}.
\end{eqnarray}
Therefore, the previous approach can be applied directly by setting  $f_{1\sigma}=f_2=u$, $f_{3\sigma}=1$ and $f_4=u(1-U\epsilon)-(u-1)\sum_{\sigma}\mu_{\sigma}\epsilon$ in Eq. (\ref{expand1}).  It immediately leads to the following partition function
\begin{eqnarray}
Z = Z_U^N\prod_{k\sigma}\left[1-Z_U^{-1}g'_{\sigma}(\omega_n)\varepsilon_{\vec{k}}\right],
\end{eqnarray}
where
\begin{eqnarray}
Z_U &=& 3+e^{-\beta(U-\sum\limits_{\sigma}\frac{u-1}{u}\mu_{\sigma})},\nonumber\\
g'_{\sigma}(\omega_n)&=&\frac{2}{i\omega_n}+\frac{\frac{1}{u}\left(1+e^{\beta(\sum\limits_{\sigma'}\frac{u-1}{u}\mu_{\sigma'}-U)}\right)}{i\omega_n-U}.\nonumber
\end{eqnarray}

\begin{figure}[ht]
\includegraphics [width=8.5cm]{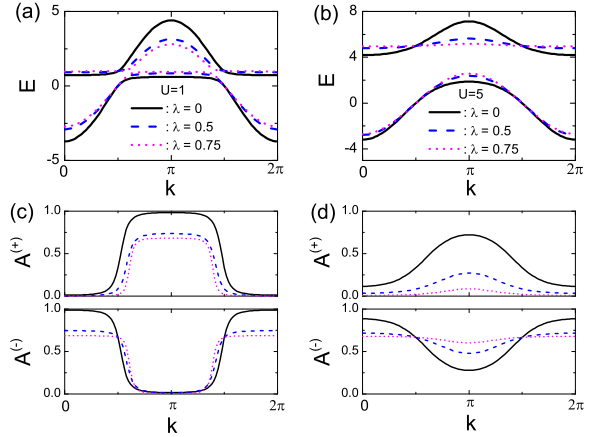}
 \caption{The Hubbard bands and corresponding spectral weights for $U=1$ (a,c) and $U=5$ (b,d), respectively. }
 \label{fig4}
\end{figure}

We here are mainly interested in the formation of the Hubbard bands at the zero temperature in the paramagnetic phase. By setting  $\mu_{\sigma}=\mu_0$ and when $\frac{u-1}{u}\mu_0<\frac{U}{2}$, we obtain the Green's function of the $\xi$ fields at the lowest temperature limit in the standard form\cite{note1}:
\begin{eqnarray}
\langle \xi_{k\sigma}^*\xi_{k\sigma}\rangle= \sum_{\nu=\pm}\frac{A^{(\nu)}_{\vec{k}\sigma}}{i\omega_n-E^{(\nu)}_{\vec{k}\sigma}}.
\end{eqnarray}
The dispersions of the quasi-particles in the upper and lower Hubbard bands are
$E^{(\pm)}_{\vec{k}\sigma}=\frac{1}{2}[U+\frac{1}{3}(2+\frac{1}{u})(\varepsilon_{\vec{k}}-\mu_0)$
$\pm\sqrt{(U+\frac{1}{3}(2+\frac{1}{u})\varepsilon_{\vec{k}}-\mu_0))^2-\frac{8}{3}(\varepsilon_{\vec{k}}-\mu_0)U}]$,  with
$ A^{(\pm)}_{\vec{k}\sigma}=\pm\frac{\frac{1}{3}(2+\frac{1}{u})E^{(\pm)}_{\vec{k}\sigma}-\frac{2}{3}U}{\sqrt{(U+\frac{1}{3}(2+\frac{1}{u})(\varepsilon_{\vec{k}}-\mu_0))^2-\frac{8}{3}(\varepsilon_{\vec{k}}-\mu_0)U}}
$ being the spectral weights. The dispersions and spectral weights are plotted Fig. \ref{fig4} for the square lattice with various $U$ and $\lambda$, fixing $t=1$ and $\mu_0=0$. We find that the band gap appears between the upper and lower Hubbard bands almost independent of $\lambda$, meaning that the Mott gap opens due to $U$ but not $\lambda$.  However, $\lambda$ influences the Hubbard bands asymmetrically because it suppresses the local double occupation states. Specifically, it significantly flattens the upper Hubbard band and reduces the corresponding spectral weight. This $\lambda$-driven correlation effect is in agreement with that exhibited by the enhanced Wilson ratio discussed previously.

Summarizing, we have developed a new coherent state path integral approach for a class of CHMs constructed in terms the deformed Hubbard operators. It allows a faithful description of the complicated hopping process in the whole region of the deformation-induced interaction parameter $\lambda$ and overcomes the divergence problem in the conventional continuous field theory approach. Interestingly, the chemical potential and the short-ranged Coulomb interaction $U$ can be also described in this approach. Our results show that the CH interaction alone always leads to the renormalization effect of the Fermi liquid even though the local double occupation states are suppressed at the Hubbard limit ($\lambda=1$) where the system locates on the verge of the correlated Fermi liquid phase. On the other hand, in the presence of finite $U$ the Mott gap opens at half-filling as usual for any $\lambda$. Increasing $\lambda$ significantly reduces the bandwidth of the upper Hubbard band and the corresponding spectral weight. These results reveal the distinct roles played by CH and Coulomb interactions, the two prototype driving forces behind the rich many-body physics. We hope that our approach can pave a way to understand the delicate interplay among various interactions in a wider family of correlated electron systems.

The authors thank C. Cao for useful discussions. This work was supported in part by the National Science Foundation of China under the grant Nos. 11874136 and 11474082.

\begin{appendix}
\section {Derivation of coherent state path-integral representation of partition function }
In this appendix, we briefly derive the path-integral representation of the partition function in terms of the coherent states of the deformed Hubbard operators.  The trace of an operator $\hat{A}$, when initially carried out in a given representation such as the energy representation in the basis of eigenstates $|n\rangle$ satisfying $\sum_{n} |n\rangle \langle n|=1 $, can be reformulated in the coherent state representation as following
\begin{eqnarray*}
&&tr\hat{A}=\sum_{n}\langle n|\hat{A}|n\rangle\\
&=&\sum_{n}\langle n|\prod_i\frac{1}{u}\int\prod_{\sigma}d\xi^*_{i\sigma}d\xi_{i\sigma}C(\xi^*_{i\sigma},\xi_{i\sigma})|\xi_i\rangle\langle\xi_i|\hat{A}|n\rangle\\
&=&\prod_i\frac{1}{u}\int\prod_{\sigma}d\xi^*_{i\sigma}d\xi_{i\sigma}C(\xi^*_{i\sigma},\xi_{i\sigma})\langle\xi_i|\hat{A}\sum_{n}|n\rangle\langle n|-\xi_i\rangle\\
&=&\prod_i\frac{1}{u}\int\prod_{\sigma}d\xi^*_{i\sigma}d\xi_{i\sigma}C(\xi^*_{i\sigma},\xi_{i\sigma})\langle\xi_i|\hat{A}|-\xi_i\rangle
\end{eqnarray*}
where the completeness relation of the coherent states of the deformed Hubbard operators (\ref{completeness}) is used, and $C(\xi^*_{i\sigma},\xi_{i\sigma})=\exp{[-u\sum\limits_{\sigma}\xi^*_{i\sigma}\xi_{i\sigma}-(u^2-u)\xi^*_{i\uparrow}\xi_{i\uparrow}\xi^*_{i\downarrow}\xi_{i\downarrow}]}$.
The minus sign in $|-\xi_i\rangle$ comes from the interchange of two Grassmann numbers.
Applying this formula to the partition function of a thermodynamic equilibrium system at temperature $T$
\begin{eqnarray}
Z = tre^{-\beta\hat{H}},
\end{eqnarray}
where $\beta=1/k_BT$ is the imaginary time interval and $k_B$ the Boltzmann constant, we have
\begin{eqnarray}\label{Z}
Z = \prod_i\frac{1}{u}\int\prod_{\sigma}d\xi^*_{i\sigma}d\xi_{i\sigma}C(\xi^*_{i\sigma},\xi_{i\sigma})\langle\xi_{i}|e^{-\beta\hat{H}}|-\xi_{i}\rangle.
\end{eqnarray}
\begin{figure}[ht]
\includegraphics [width=6cm]{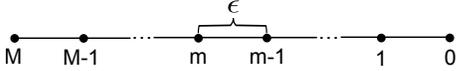}
 \caption{The imaginary time slices with width $\epsilon=\beta/M$.} \label{fig5}
\end{figure}

Dividing $\beta$ by M number of slices with equal width $\epsilon=\beta/M$ using $e^{-\beta \hat{H}}= (e^{-\epsilon  \hat{H}})^M $ and inserting the completeness relation between each operator $e^{-\epsilon\hat{H}}$, we have
\begin{eqnarray*}
Z&=&\int\prod_{i}\prod_{m=1}^{M}\frac{1}{u}\int\prod_{\sigma}d\xi^*_{i\sigma}d\xi_{i\sigma}C(\xi^*_{i\sigma,m}\xi_{i\sigma,m})\\
&&\langle\xi_{i,M}|e^{-\epsilon\hat{H}}|\xi_{i,M-1}\rangle\langle\xi_{i,M-1}|e^{-\epsilon\hat{H}}|\xi_{i,M-2}\rangle\\
&&\cdots\langle\xi_{i,m}|e^{-\epsilon\hat{H}}|\xi_{i,m-1}\rangle\cdots\\
&&\langle\xi_{i,2}|e^{-\epsilon\hat{H}}|\xi_{i,1}\rangle\langle\xi_{i,1}|e^{-\epsilon\hat{H}}|-\xi_{i,M}\rangle,
\end{eqnarray*}
where $m$ denotes the location of each inserts, playing the role of the discrete imaginary time. The $\xi$ fields in Eq.(\ref{Z}) is denoted by $\xi_{i\sigma,M}$.

In the limit $\epsilon\rightarrow 0 $, the matrix element within neighboring time slices is given by
\begin{eqnarray*}
&&\langle\xi_{i,m}|e^{-\epsilon\hat{H}}|\xi_{i,m-1}\rangle\\
&=&\langle\xi_{i,m}|1-\epsilon\hat{H}|\xi_{i,m-1}\rangle\\
&=& e^{-\epsilon H(\xi^*_{i\sigma,m},\xi_{i\sigma,m-1})}\langle\xi_{i,m}|\xi_{i,m-1}\rangle \\
&=&e^{-\epsilon H(\xi^*_{i\sigma,m},\xi_{i\sigma,m-1})+\sum\limits_{\sigma}\xi^*_{i\sigma,m}\xi_{i\sigma,m-1}+(u-1)\prod\limits_{\sigma}\xi^*_{i\sigma,m}\xi_{i\sigma,m-1}}.
\end{eqnarray*}
Here, the Hamiltonian is normal ordered with the element $ H(\xi^*_{i\sigma,m},\xi_{i\sigma,m-1})$.
Therefore, taking the anti-periodic boundary condition $\xi_{i\sigma,0} = -\xi_{i\sigma,M}$, the partition function can be expressed in the following path-integral
\begin{equation}
Z = \int D(\xi^*,\xi)e^{-S(\xi^*,\xi)},
\end{equation}
where $D(\xi^*,\xi)= \lim\limits_{M\rightarrow\infty}\prod\limits_{i,m}(\frac{1}{u}) d\xi^*_{i\uparrow,m}d\xi_{i\uparrow,m}d\xi^*_{i\downarrow,m}d\xi_{i\downarrow,m}$ and
\begin{eqnarray}
S &=&\sum_{i,m,\sigma}(u\tilde{n}_{i\sigma,m}-\xi^*_{i\sigma,m}\xi_{i\sigma,m-1})\nonumber\\
&+&\sum_{i,m}(u^2-u)\prod_{\sigma}\tilde{n}_{i\sigma,m}-\sum_{i,m}(u-1)\prod_{\sigma}\xi^*_{i\sigma,m}\xi_{i\sigma,m-1}\nonumber\\
&+&\sum_{m}\epsilon H[\{\xi^*_{m}\},\{\xi_{m-1}\}],
\end{eqnarray}
with $\tilde{n}_{i\sigma,m}=\xi^*_{i\sigma,m}\xi_{i\sigma,m}$.

\section{The Properties of Grassmann integrations}
The most useful Grassmann  integrations used in the present path-integral approach are listed below:
\begin{eqnarray}
&&\int d\xi^*d\xi e^{-h\xi^*\xi}=h,\\
&&\int \prod_{\sigma}d\xi_{\sigma}^*d\xi_{\sigma} e^{-U\xi_{\uparrow}^*\xi_{\uparrow}\xi_{\downarrow}^*\xi_{\downarrow}}=-U,
\end{eqnarray}

\begin{eqnarray}
&&\int \prod_{m=1}^Md\xi_m^*d\xi_m \prod_{m=1}^M\xi^*_{m}\xi_{m-1}\nonumber\\
&=&\int \prod_{m=1}^Md\xi_m^*d\xi_m \left(\xi^*_{M}\xi_{M-1}\cdots\xi_2^*\xi_1\xi_1^*(-\xi_M)\right)\nonumber\\
&=&\int \prod_{m=1}^Md\xi_m^*d\xi_m \left(\xi_M\xi^*_{M}\xi_{M-1}\cdots\xi_2^*\xi_1\xi_1^*\right)=1,
\end{eqnarray}
where the anti-periodic boundary condition $\xi_0=-\xi_M$ is imposed.

\section{the partition function for the single site toy model}
Here we exactly solve the single site problem using the conventional  particle number representation. The Hamiltonian of this toy model is
\begin{eqnarray}\label{Hu}
\hat{H}&=&-\sum_{\sigma}w_{\sigma}(\tilde{X}^{0\sigma})^{\dag}\tilde{X}^{0\sigma}\nonumber\\
&=&-\sum_{\sigma}w_{\sigma}(1-\lambda \hat{n}_{\bar{\sigma}})\hat{n}_{\sigma}(1-\lambda \hat{n}_{\bar{\sigma}})\nonumber\\
&=&-\sum_{\sigma}w_{\sigma}\left[\hat{n}_{\sigma}+(\lambda^2-2\lambda)\hat{n}_{\uparrow}\hat{n}_{\downarrow}\right]
\end{eqnarray}
It is diagonal in the particle number representation with eigenvalues $E_1=0$, $E_2=-w_{\uparrow}$,  $E_3=-w_{\downarrow}$ and $E_4=-\sum\limits_{\sigma}(1-\lambda)^2w_{\sigma}$ respectively. Therefore, we have
\begin{eqnarray}
Z=tre^{-\beta\hat{H}}=\sum_{i=1}^4e^{-\beta E_i}=1+e^{\beta\sum\limits_{\sigma}\frac{w_{\sigma}}{u}}+\sum_{\sigma}e^{\beta w_{\sigma}}
\end{eqnarray}
where $u=(1-\lambda)^{-2}$.

\section{Derivation of $g_{\sigma}(\omega_n)$ in Eq. (\ref{g})}
The contribution from diagrams illustrated in Fig. \ref{fig2} (a) is given by
 \begin{eqnarray}\label{a}
&&\epsilon^2\sum_{ l=0}^{M-1}\sum_{\sigma,m=1}^M(1+\epsilon w_{\sigma})^{l}\eta^*_{i\sigma,m+l}\eta_{i\sigma,m}\nonumber\\
&=&\epsilon\sum_{l=0}^{M-1}\sum_{\sigma,\omega_n}(1+\epsilon w_{\sigma})^{l}e^{i\omega_n\epsilon l}\eta^*_{i\sigma}(\omega_n)\eta_{i\sigma}(\omega_n)\nonumber\\
&=&\epsilon\sum_{\sigma,\omega_n}\frac{1-(1+\epsilon w_{\sigma})^{M}e^{i\omega_n\epsilon M}}{1-(1+\epsilon w_{\sigma})e^{i\omega_n\epsilon}}\eta^*_{i\sigma}(\omega_n)\eta_{i\sigma}(\omega_n),
\end{eqnarray}
where the Fourier transformation for the $\eta$ fields is applied
\begin{eqnarray}
\eta_{i\sigma,m}&=&\frac{1}{\sqrt{\beta}}\sum_{\omega_n}\eta_{i\sigma}(\omega_n)e^{-i\omega_n\tau_m}.
\end{eqnarray}
Since $e^{i\omega_n\epsilon M}=-1$ and $\lim_{M\rightarrow\infty}(1+\epsilon w_{\sigma})^{M}$=$e^{\beta w_{\sigma}}$, Eq.(\ref{a}) becomes
\begin{eqnarray}\label{g1}
-\sum_{\sigma,\omega_n}\frac{1+e^{\beta w_{\sigma}}}{i\omega_n+w_{\sigma}}\eta^*_{i\sigma}(\omega_n)\eta_{i\sigma}(\omega_n).
\end{eqnarray}

The contribution from diagrams illustrated in Fig. \ref{fig2} (b) is given by
 \begin{eqnarray}
\epsilon^2\sum_{\sigma,l=0}^{M-1}\sum_{m=1}^M\frac{(1+\epsilon w_{\bar{\sigma}})^{M}}{u}\left(\frac{1+\epsilon\sum\limits_{\sigma'}\frac{w_{\sigma'}}{u}}{1+\epsilon w_{\bar{\sigma}}}\right)^l\eta^*_{i\sigma,m+l}\eta_{i\sigma,m}.\nonumber
\end{eqnarray}
Taken summation over $m$, it becomes
 \begin{eqnarray}
\epsilon\sum_{\sigma,l=0}^{M-1}\sum_{\omega_n}\frac{e^{\beta w_{\bar{\sigma}}}}{u}\left(\frac{1+\epsilon\sum\limits_{\sigma'}\frac{w_{\sigma'}}{u}}{1+\epsilon w_{\bar{\sigma}}}\right)^le^{i\omega_n\epsilon l}\eta^*_{i\sigma}(\omega_n)\eta_{i\sigma}(\omega_n).\nonumber
\end{eqnarray}
Again, taken summation over $l$, it becomes
 \begin{eqnarray}\label{g2}
-\sum_{\sigma,\omega_n}\frac{\frac{1}{u}\left(e^{\beta w_{\bar{\sigma}}}+e^{\beta\sum\limits_{\sigma'}\frac{w_{\sigma'}}{u}}\right)}{i\omega_n+\sum\limits_{\sigma'}\frac{w_{\sigma'}}{u}-w_{\bar{\sigma}}}\eta^*_{i\sigma}(\omega_n)\eta_{i\sigma}(\omega_n)
\end{eqnarray}

The sum of Eq. (\ref{g1}) and Eq. (\ref{g2}) leads to the contributions from the quadratic term of the $\xi$ fields in expanding the exponential of expression (\ref{St}), namely, $-\sum\limits_{\omega_n,\sigma}g_{\sigma}(\omega_n)\eta^*_{i\sigma}(\omega_n)\eta_{i\sigma}(\omega_n)$ , with
\begin{eqnarray}
g_{\sigma}(\omega_n)=\frac{1+e^{\beta w_{\sigma}}}{i\omega_n+w_{\sigma}}+\frac{\frac{1}{u}\left(e^{\beta w_{\bar{\sigma}}}+e^{\beta\sum_{\sigma'}\frac{w_{\sigma'}}{u}}\right)}{i\omega_n+\sum_{\sigma'}\frac{w_{\sigma'}}{u}-w_{\bar{\sigma}}}.
\end{eqnarray}

\section{Magnetization and Susceptibility for CHM}
According to Eq. (\ref{Hu}),
 \begin{eqnarray}
 \hat{n}_{i\uparrow}-\hat{n}_{i\downarrow} = (\tilde{X}^{0\uparrow})^{\dag}\tilde{X}^{0\uparrow}-(\tilde{X}^{0\downarrow})^{\dag}\tilde{X}^{0\downarrow}.
  \end{eqnarray}
This operator is commutative with the CH term defined in the Hamiltonian (\ref{H}).
By choosing $w_{\sigma}=\sigma h$ with $h$ being the external magnetic field and $\sigma=1$ for up spin and $\sigma=-1$ for down spin, the magnetization density is given by
\begin{eqnarray}
m =\frac{1}{N\beta}\frac{\partial\ln{Z}}{\partial h}.
\end{eqnarray}
The partition function $Z$ is given by Eq. (\ref{Hc}), where $Z_w=(1+e^{\beta\sigma h})(1+e^{-\beta\sigma h})$ and $Z_w^{-1}g_{\sigma}(\omega_n)\varepsilon_{\vec{k}}=\frac{1+\frac{1}{u}e^{-\beta\sigma h}}{1+e^{-\beta\sigma h}}\frac{\varepsilon_{\vec{k}}}{i\omega_n+\sigma h}$. Therefore, we have
\begin{eqnarray}
m& =&\frac{\sigma (e^{\beta\sigma h}-e^{-\beta\sigma h})}{Z_w}-\frac{1}{N\beta}\sum_{k\sigma}\frac{\sigma}{i\omega_n+\sigma h}\nonumber\\
&&+\frac{1}{N\beta}\sum_{k\sigma}\frac{\sigma-\sigma\frac{(1-\frac{1}{u})}{Z_w}\beta\varepsilon_{\vec{k}}}{i\omega_n+\sigma h - \frac{1+\frac{1}{u}e^{-\beta\sigma h}}{1+e^{-\beta\sigma h}}\varepsilon_{\vec{k}}}\nonumber\\
&=&\frac{1}{N}\sum_{\vec{k}\sigma}\sigma(1-\frac{1-\frac{1}{u}}{Z_1}\beta\varepsilon_{\vec{k}}) F(-\sigma h + \frac{1+\frac{1}{u}e^{-\beta\sigma h}}{1+e^{-\beta\sigma h}}\varepsilon_{\vec{k}}), \nonumber
\end{eqnarray}
where $F(x)=\frac{1}{1+e^{\beta x}}$ is the Fermi-Dirac distribution function.
The susceptibility $\chi = \frac{\partial m}{\partial h}|_{h=0}$ is then given by
\begin{eqnarray}
\frac{\beta}{N}\sum_{\vec{k}\sigma}(1 - \beta\varepsilon_{\vec{k}}\frac{1-\frac{1}{u}}{4})^2F( \varepsilon_{\vec{k}}\frac{1+\frac{1}{u}}{2})[1-F( \varepsilon_{\vec{k}}\frac{1+\frac{1}{u}}{2})].\nonumber
\end{eqnarray}

This summation can be represented by the integral over the energy by introducing the density of state $N(\varepsilon)$ for the free fermions
\begin{eqnarray}
\chi &=& 2\beta\int_{-\Lambda}^{\Lambda}d\varepsilon N(\varepsilon)\left(1 -2 \beta\varepsilon\frac{1-\frac{1}{u}}{4}+(\beta\varepsilon\frac{1-\frac{1}{u}}{4})^2\right)\nonumber\\
&&F( \varepsilon\frac{1+\frac{1}{u}}{2})[1-F( \varepsilon\frac{1+\frac{1}{u}}{2})]\nonumber\\
&=&\frac{4}{1+\frac{1}{u}}N(0)+\frac{\pi^2}{3}\frac{(1-\frac{1}{u})^2}{(1+\frac{1}{u})^3} N(0),
\end{eqnarray}
 with the $2\Lambda$ being the bandwidth of the free electrons. In above, we have set $\mu_B=1$ and used the $\delta$-function representation
\begin{eqnarray}
\frac{\beta}{(e^{\beta x}+1)(e^{-\beta x}+1)}=\delta(x)
\end{eqnarray}
which is valid in the low temperature limit $\beta\rightarrow \infty$. The integration range of $\beta\varepsilon$  tends to $\infty$ in this limit, too, leading to
\begin{eqnarray}
\int_{-\infty}^{\infty}\frac{x^2dx}{(e^x+1)(e^{-x}+1)}=\frac{\pi^2}{3}.
\end{eqnarray}

\section{Internal energy and specific heat for CHM}
In order to calculate the internal energy, \begin{eqnarray}
E = -\frac{\partial \ln Z}{\partial\beta},
\end{eqnarray}
we can set $\mu_{\sigma}=0$ so that $Z_w=4$ and $Z_w^{-1}g_{\sigma}(\omega_n)=\frac{1}{2}(1+\frac{1}{u})\frac{1}{i\omega_n}$.  Then we have
\begin{eqnarray}
\ln{Z}&=&N\ln{Z_w}+\sum_{k\sigma}\ln{[1-Z_w^{-1}g_{\sigma}(\omega_n)\varepsilon_{\vec{k}}]}\nonumber\\
&=&N\ln{Z_w}+\sum_{k\sigma}\ln{[i\omega_n-\frac{1}{2}(1+\frac{1}{u})\varepsilon_{\vec{k}}]}-\ln{(i\omega_n)}.\nonumber
\end{eqnarray}
Introducing a temporary variable $\alpha $ such that
\begin{eqnarray}
\ln{Z_{\alpha}}&=&N\ln{Z_w}+\nonumber\\
&+&\sum_{k\sigma}\ln{[i\omega_n-\alpha-\frac{1}{2}(1+\frac{1}{u})\varepsilon_{\vec{k}}]}-\ln{(i\omega_n-\alpha)}.\nonumber
\end{eqnarray}
The internal energy can be reexpressed as
\begin{eqnarray}
E = -\left.\left(\frac{\partial}{\partial\beta}\int d\alpha \frac{\partial}{\partial\alpha}\ln Z_{\alpha}\right)\right|_{\alpha=0}.\nonumber
\end{eqnarray}
Because
\begin{eqnarray}
&&-\frac{\partial}{\partial\beta}\int d\alpha \frac{\partial}{\partial\alpha}\ln Z_{\alpha}\nonumber\\
&=&\frac{\partial}{\partial\beta}\int d\alpha \sum_{k\sigma}\left(\frac{1}{i\omega_n-\alpha-\frac{1}{2}(1+\frac{1}{u})\varepsilon_{\vec{k}}}-\frac{1}{i\omega_n-\alpha}\right)\nonumber\\
&=&\frac{\partial}{\partial\beta}\int d(\beta\alpha) \sum_{\vec{k}\sigma}\left[F\left(\alpha+\frac{1}{2}(1+\frac{1}{u})\varepsilon_{\vec{k}}\right)-F\left(\alpha\right)\right],\nonumber
\end{eqnarray}
we have
\begin{eqnarray}
E = \sum_{\vec{k}}(1+\frac{1}{u})\varepsilon_{\vec{k}}F\left(\frac{1}{2}(1+\frac{1}{u})\varepsilon_{\vec{k}}\right).
\end{eqnarray}
Then the specific heat can be derived as in the following:
\begin{eqnarray}
C&=& \frac{\partial E}{\partial T}= -k_B\beta^2\frac{\partial E}{\partial \beta}\nonumber\\
&=&2k_B\beta^2\sum_{\vec{k}}\left(\frac{1}{2}(1+\frac{1}{u})\varepsilon_{\vec{k}}\right)^2\nonumber\\
&&F\left(\frac{1}{2}(1+\frac{1}{u})\varepsilon_k\right)\left[1-F\left(\frac{1}{2}(1+\frac{1}{u})\varepsilon_{\vec{k}}\right)\right]\nonumber\\
&=&2k_B\beta^2\int_{-\Lambda}^{\Lambda} d\varepsilon N(\varepsilon)\left(\frac{1}{2}(1+\frac{1}{u})\varepsilon\right)^2\nonumber\\
&&F\left(\frac{1}{2}(1+\frac{1}{u})\varepsilon\right)\left[1-F\left(\frac{1}{2}(1+\frac{1}{u})\varepsilon\right)\right]\nonumber\\
&=&\frac{4}{1+\frac{1}{u}}N(0)k_B^2T.
\end{eqnarray}

\section{The Green's function of $\xi$ fields}

To calculate the Green's function of the $\xi$ fields, we introduce the corresponding external sources by adding $\sum_{k\sigma}(J^*_{k\sigma}\xi_{k\sigma}+\xi_{k\sigma}^*J_{k\sigma})$ to $S_t$ in Eq. (\ref{St}). Then, the partition function becomes
\begin{eqnarray}
Z = Z_U^N \prod_{k\sigma}[1-Z_U^{-1}g'_{\sigma}(\omega_n)\varepsilon_{\vec{k}}-Z_U^{-1}g'_{\sigma}(\omega_n)J^*_{k\sigma}J_{k\sigma}].
\end{eqnarray}
It leads to
\begin{eqnarray}
\langle \xi_{k\sigma}^*\xi_{k\sigma}\rangle= -\left.\frac{\partial^2\ln Z}{\partial J_{k\sigma}\partial J^*_{k\sigma}}\right|_{J=0}=\frac{Z_U^{-1}g'_{\sigma}(\omega_n)}{1-Z_U^{-1}g'_{\sigma}(\omega_n)\varepsilon_{\vec{k}}}.
\end{eqnarray}
This can be expressed in the following standard form
\begin{eqnarray}
\langle \xi_{k\sigma}^*\xi_{k\sigma}\rangle= \sum_{\nu=\pm}\frac{A^{(\nu)}_{\vec{k}\sigma}}{i\omega_n-E^{(\nu)}_{\vec{k}\sigma}}.
\end{eqnarray}

By taking the zero temperature limit and setting $\mu_{\sigma}=\mu_0$, we have
\begin{eqnarray}
E^{(\pm)}_{\vec{k}\sigma}&=&\frac{1}{2}\left(U+\frac{2+\frac{1}{u}}{3}(\varepsilon_{\vec{k}}-\mu_0)\right.\nonumber\\
&\pm&\left.\sqrt{(U+\frac{2+\frac{1}{u}}{3}(\varepsilon_{\vec{k}}-\mu_0))^2-\frac{8}{3}(\varepsilon_{\vec{k}}-\mu_0)U}\right).
\end{eqnarray}
These are the dispersions of the quasi-particles in the upper and lower Hubbard bands respectively,  with the spectral weights
\begin{eqnarray}
A^{(\pm)}_{\vec{k}\sigma}=\pm\frac{\frac{2+\frac{1}{u}}{3}E^{(\pm)}_{\vec{k}\sigma}-\frac{2}{3}U}{\sqrt{(U+\frac{2+\frac{1}{u}}{3}(\varepsilon_{\vec{k}}-\mu_0))^2-\frac{8}{3}(\varepsilon_{\vec{k}}-\mu_0)U}}
\end{eqnarray}
satisfying $A^{(\pm)}_{\vec{k}\sigma}>0$ and $\sum\limits_{\nu=\pm}A^{(\pm)}_{\vec{k}\sigma}=\frac{1}{3}(2+\frac{1}{u})$.

\section{Chemical potential and Hubbard interaction expressed in terms of deformed Hubbard operators}
The on-site Coulomb interaction can be expressed as
\begin{eqnarray}\label{onsite}
\hat{n}_{\uparrow}\hat{n}_{\downarrow}=u(\tilde{X}^{0\downarrow})^{\dag}(\tilde{X}^{0\uparrow})^{\dag}\tilde{X}^{0\uparrow}\tilde{X}^{0\downarrow}.
\end{eqnarray}
where $u=(1-\lambda)^{-1}$.
In order to investigate the generic case with tunable electron filling, we express the particle number operators $\hat{n}_{\sigma}$ in terms of the deformed Hubbard operators.  According to Eq. (\ref{Hu})
\begin{eqnarray}
(\tilde{X}^{0\sigma})^{\dag}\tilde{X}^{0\sigma}=\hat{n}_{\sigma}+(\lambda^2-2\lambda)\hat{n}_{\uparrow}\hat{n}_{\downarrow},
\end{eqnarray}
and combining it with Eq.(\ref{onsite}), we obtain the following expression for the particle number operators
\begin{eqnarray}
\hat{n}_{\sigma} &=& (\tilde{X}^{0\sigma})^{\dag}\tilde{X}^{0\sigma}\nonumber\\
&+&(u-1)(\tilde{X}^{0\downarrow})^{\dag}(\tilde{X}^{0\uparrow})^{\dag}\tilde{X}^{0\uparrow}\tilde{X}^{0\downarrow}.
\end{eqnarray}

\end{appendix}


\begin{thebibliography}{99}
\bibitem{note1} More details about the path-integral representation and other related derivations discussed in the main context are provided in Appendices A-H.
\end{thebibliography}
\end{document}